\documentclass[letterpaper, 10 pt, conference]{ieeeconf}  %

\IEEEoverridecommandlockouts                              %

\usepackage{graphicx}
\usepackage{xcolor}

\title{\LARGE \bf
Identifying shifts in multi-modal travel patterns during special events using mobile data: Celebrating Vappu in Helsinki
}

\author{Zhiren Huang$^{1}$, Charalampos Sipetas$^{2}$, Alonso Espinosa Mireles de Villafranca$^{2}$ and Tri Quach$^{3}$%
\thanks{$^{1}$Zhiren Huang is with Department of Computer Science, Aalto University, 02150 Espoo, Finland
        {\tt\small zhiren.huang@aalto.fi}}%
\thanks{$^{2}$Charalampos Sipetas and Alonso Espinosa Mireles de Villafranca are with the Department of Built Environment, Aalto University, 02150 Espoo, Finland
        }%
\thanks{$^{3}$Tri Quach is with Department of Analytics and Research, HSL Helsingin seudun liikenne (Helsinki Regional Transport Authority), 00520 Helsinki, Finland
        }%
}
\begin{document}

\maketitle
\thispagestyle{empty}
\pagestyle{empty}

\begin{abstract}

Large urban special events significantly contribute to a city's vibrancy and economic growth but concurrently impose challenges on transportation systems due to alterations in mobility patterns. This study aims to shed light on mobility patterns by utilizing a unique, comprehensive dataset collected from the Helsinki public transport mobile application and Bluetooth beacons. Earlier methods, relying on mobile phone records or focusing on single traffic modes, do not fully grasp the intricacies of travel behavior during such events. We focus on the Vappu festivities (May 1st) in the Helsinki Metropolitan Area, a national holiday characterized by mass gatherings and outdoor activities. We examine and compare multi-modal mobility patterns during the event with those during typical non-working days in May 2022. Through this case study, we find that people tend to favor public transport over private cars and are prepared to walk longer distances to participate in the event. The study underscores the value of using comprehensive multi-modal data to better understand and manage transportation during large-scale events.

\end{abstract}

\begin{keywords}
Public transport, Crowded events, Human mobility, Mobile ticketing application, Bluetooth beacons
\end{keywords}

\section{Introduction}

Large-scale special events are crucial for a city's vibrancy \cite{McGillivray2019} and economic growth \cite{Richards2015}, but they also present significant challenges to transportation systems due to the complex mobility patterns they generate \cite{Parkes2016}. For example, people's daily routine may change \cite{Guo2022}, leading to changes in travel mode choice and the overall travel demand. Without efficient crowd management \cite{Johansson2012} and public traffic coordination \cite{Xu2017}, such changes can lead to severe congestion \cite{Humphreys2018, Wang2022} or even fatal stampedes \cite{Helbing2007, Pretorius2013}, such as the Seoul Halloween crowd crush in 2022 \cite{Sharma2023}. Hence, understanding people's travel demand and behaviours during crowded events is of major importance.

For understanding crowd behaviour, video surveillance has been widely applied to study pedestrian movement patterns within event venues \cite{Helbing2007} and to determine safe crowd density thresholds (i.e., how many people per square meter) \cite{Pretorius2013}. However, those studies focus on site-specific features, such as the exits or corridors. With the rapid development of Information and Communications Technology (ICT), sensing large-scale crowd movements has become feasible \cite{Kaiser2018}. Mobile phone data starts to play an important role in sensing large-scale crowd gathering processes. Candia et al. \cite{Candia2008} and Dong et al. \cite{Dong2015} use Call Detail Record (CDR) data to detect unusual events in cities. Fan et al \cite{Fan2015} use mobile phone GPS logs to predict residents' short-term future movements during crowded events in Tokyo. Xu and González \cite{Xu2017} combine multi-source data, such as CDR, transit schedules and Airbnb data to analyze the travel patterns during the Olympic Games in Rio de Janeiro. Zhou et al. \cite{Zhou2017} use online map query data to demonstrate the potential of using mobile map application for early warning of the Shanghai Bund stampede. Nevertheless, the common limitation of mobile phone data is the lack of comprehensive travel mode information during such events. Other data sources include traffic data collected from intelligent transportation systems (ITS), such as smart card data. Campanella et al. \cite{Campanella2013} use the metro operational data (i.e., ridership) to systemically analyze one metro station near the beach of Copacabana in Rio de Janeiro during New Years Eve celebration. Guo et al. \cite{Guo2022} use metro smart card data to study individual and collective behaviours during 12 crowded events in Shenzhen, China. Huang et al. \cite{Huang2018} combine bus and subway smart card data, and taxi GPS to identify the essential mobility network structures of urban crowd gathering processes. Those approaches, however, are limited to specific traffic modes. Therefore, how people shifted their travel mode choice during the respective crowded event is unknown. For example, how many people use public transport (PT) to arrive at critical locations related with a special event could be used for PT operation planning. Hence, crowd management \cite{Johansson2012, Batty2012} and evacuation design \cite{Klüpfel2013, Zhou2021, Gkiotsalitis2022} could benefit from travel mode information concerning crowded events and special celebrations.

To tackle the obscure travel mode problem, this study presents the spatio-temporal analysis of multi-modal mobility patterns during special events by leveraging an emerging data source. More specifically, "TravelSense" is a pilot project of the Helsinki Regional Transport Authority (HSL) that uses Bluetooth beacons installed in PT vehicles combined with a mobile phone ticket app to collect door-to-door trajectories from anonymous PT users \cite{Huang2022}. The generated dataset allows for a more accurate and nuanced understanding of mobility patterns during large-scale events, such as the May 1st, or \emph{Vappu}, celebration in the Helsinki Metropolitan Area.

The main contributions of this study are: 
\begin{itemize}
    \item Demonstrating the usage of real world data, collected from a mobile PT application and Bluetooth beacons, to study the multi-modal mobility patterns during special events.
    \item Understanding mode shifts and changes in travel behavior during special days, also through comparison with other non-working days (i.e., weekends and Ascension Day that is also a holiday in Helsinki during May).
    \item Quantifying the role of PT in coordinating people to attend special events in an urban area.
\end{itemize}
Such contributions pave the way for studies focusing on how to coordinate PT operation and management for large urban activities and crowded events, and they could further lead to the development of ITS that account for the complexities of large-scale urban events.

\section{Materials and Framework} 

\subsection{Data Sources}
\begin{figure}[tpb]
  \centering
  \includegraphics[width=8.5cm]{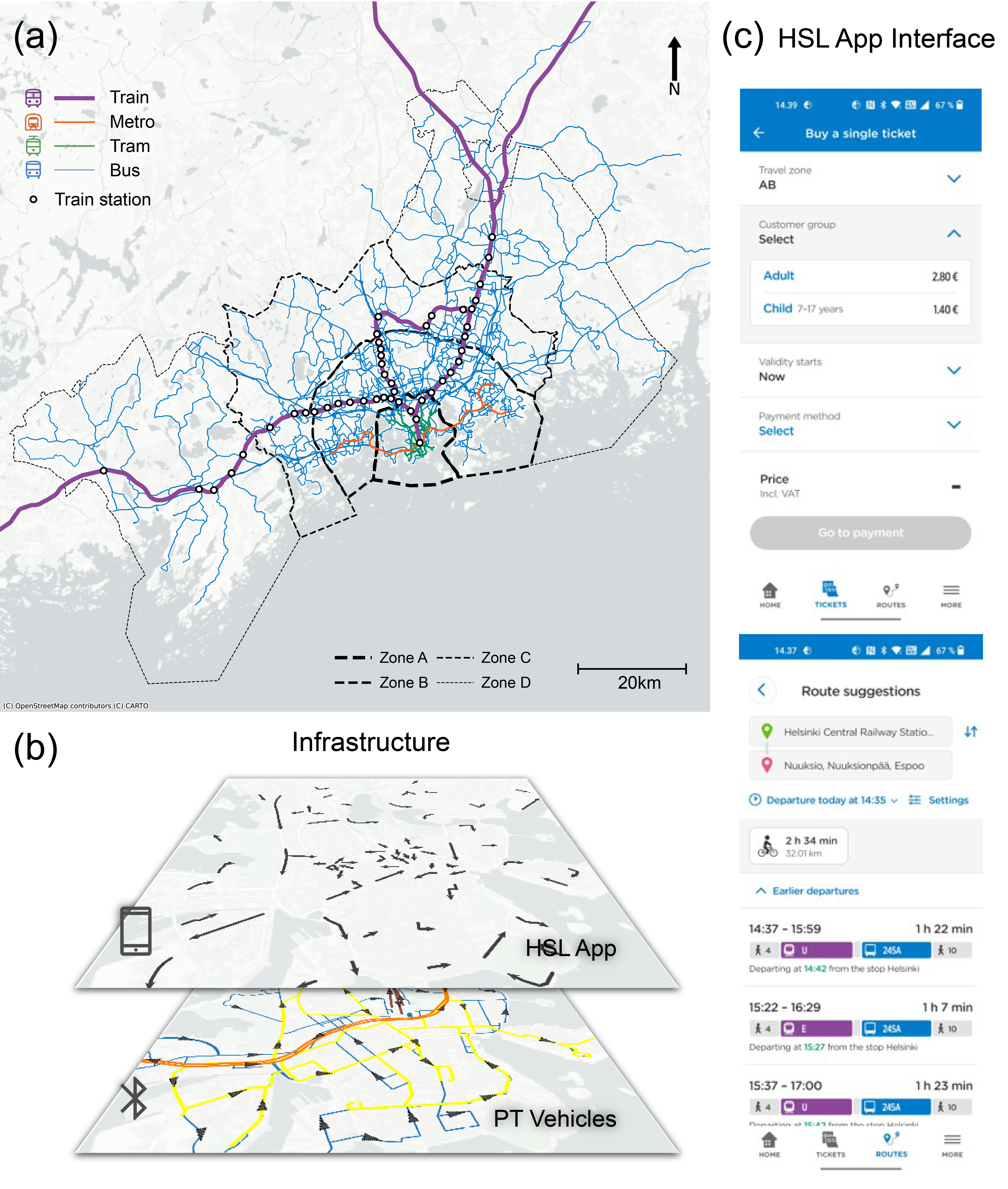}
  \caption{Data and study area overview. (a) PT network of Helsinki metropolitan area; (b) Infrastructure of TravelSense; (c) Screenshots of HSL mobile ticket application.}
  \label{fig:data}
\end{figure}

This study utilizes data collected from the 1st to the 31st of May 2022, obtained from HSL's TravelSense pilot project. HSL provides PT services for around 1.2 million residents across nine municipalities in Helsinki metropolitan area. The study area and the PT network are illustrated in Fig.\ref{fig:data}a. To facilitate PT trips, HSL offers a mobile phone ticketing application that enables passengers to easily buy tickets, search the best routes, and receive timely updates about PT operations (Fig.\ref{fig:data}c). The data collection infrastructure (Fig.~\ref{fig:data}b) incorporates two components, the first employs the HSL mobile ticketing application to sense non-PT trips, while the other utilizes Bluetooth beacons (Fig.\ref{fig:data}b) installed in PT vehicles (including subway, train, tram, bus, and ferry) to capture PT trips.  By integrating data from these infrastructures, we acquire door-to-door trajectories of anonymised passengers. To ensure high data privacy standards, devices are anonymised daily with random IDs, and location coordinates are only accurate to grid cells of 250m x 250m. Additionally, location timestamps outside the PT network are rounded to the nearest quarter-hour.

The data prepossessing steps, as described in \cite{Huang2022}, involved cleaning the raw data and aggregating the data at different levels. The information is structured based on \emph{legs} (i.e., trips) and \emph{trip chains} (i.e., journeys or trajectories). A trip chain is characterised by the legs that make up each segment and the travel modes used in each of them \cite{Raty2018}. The accuracy of mobile phone activity detection was proved in \cite{Rinne2017} and the representativeness of TravelSense is partly validated by external data sources: in \cite{Huang2022} with mobile phone trips data and in \cite{Huang2023} with commuter train automatic passenger count (APC) data.

\subsection{Event Background}

\begin{figure}[tpb]
  \centering
  \includegraphics[width=8.5cm]{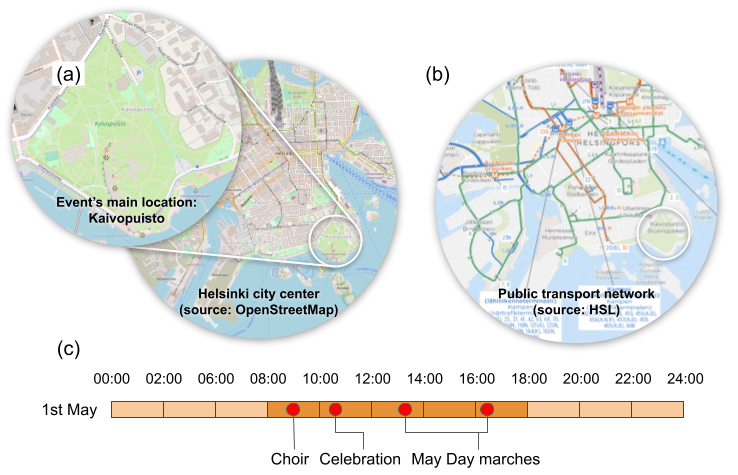}
  \caption{Event context. (a) Main location - Kaivopuisto; (b) PT near Kaivopuisto; (c) timeline of Vappu 2022.} 
  \label{fig:event}
\end{figure}

The study focuses on the \emph{Vappu} celebration of May 1st in Helsinki, Finland. There are several popular locations around Helsinki during the Vappu celebration, with the biggest being Kaivopuisto, a park area of approximately 0.45 $\textrm{km}^2$ (Fig.\ref{fig:event}(a)). Traditionally, people gather together in this park for several recreational activities throughout the day. The PT operator (i.e., HSL) offers modified services during that day, which are shown in Fig.\ref{fig:event}(b). Regarding the event's time schedule, the main part of celebrations starts at 8 a.m. and lasts until 6 p.m. (Fig.\ref{fig:event}(c)). The detailed report and timeline of 2022 Vappu event can be found in the news of the Finnish Broadcasting Company (YLE) \cite{Yle2022}, including also visual proof of the extraordinary crowding levels.

\subsection{Framework}

This study refers to the Vappu celebration of May 1st 2022 which happened to be a Sunday and includes TravelSense data from the entire metropolitan area of Helsinki. For comparison, data from other non-working days are also utilized, including  four weekends and one National holiday (i.e., Ascension day) during May 2022. 

\begin{figure}[tpb]
  \centering
  \includegraphics[width=8.5cm]{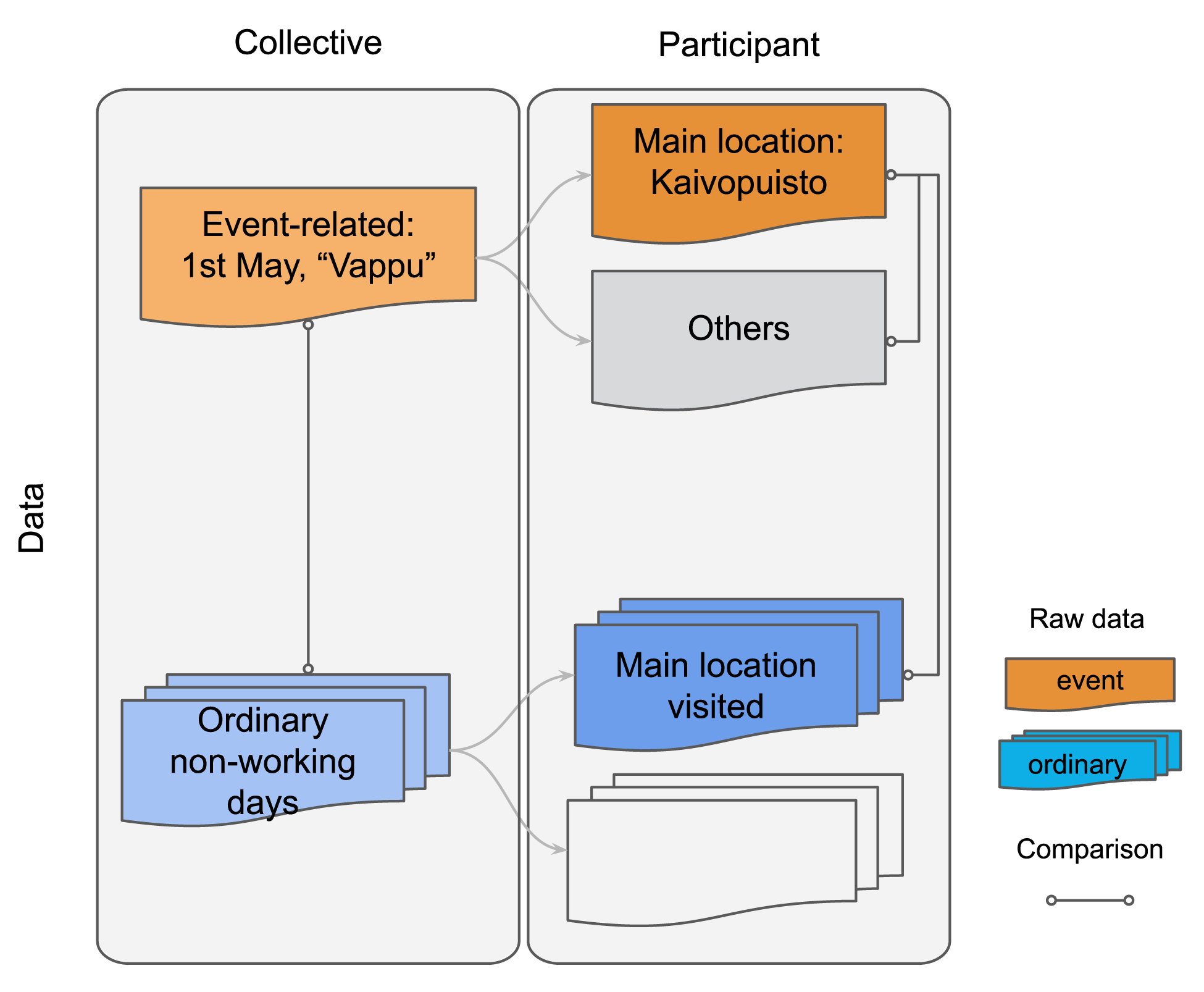}
  \caption{Analysis framework.}
  \label{fig:framework}
\end{figure}

Fig.\ref{fig:framework} presents the analysis framework for understanding the multi-modal mobility patterns during this special event. As shown, a combination of descriptive statistics, spatial analysis techniques, and data visualization tools are employed for deriving the results and findings of this study. More specifically, the analysis focused on the following aspects:

\begin{itemize}
    \item Collective level: We compared the overall spatial and temporal distribution of trajectories during the Vappu celebration and ordinary non-working days (i.e., ordinary weekends), since Vappu was a Sunday in the year of study.  Furthermore, we examined the variations in travel mode preferences. We utilised heatmaps and t-tests to quantify the differences in mobility patterns, providing insights into collective shifts in transport usage. These insights are invaluable for traffic and transport planners in managing such events.
    \item Participant level: We focused on travel behavior characteristics of the travelers that access the main event location (i.e., Kaivopuisto park). This involves assessing variations in travel distances, modes of transport, and the proportion of walking during the Vappu event in comparison to ordinary non-working days. Through this analysis, we can highlight unique characteristics in mobility patterns during special events, demonstrating how participants adjust their behavior to accommodate the event's demand.
\end{itemize}

\section{Results}

\subsection{Collective patterns during the event}

In alignment with previous studies on special events (\cite{Candia2008, Dong2015,Huang2018}), our analysis begins with examining general patterns from spatial and temporal perspectives, as well as the composition of travel modes. The analysis in this part refers to the collective level (see Fig.\ref{fig:framework}).

\begin{figure}[tpb]
  \centering
  \includegraphics[width=8.5cm]{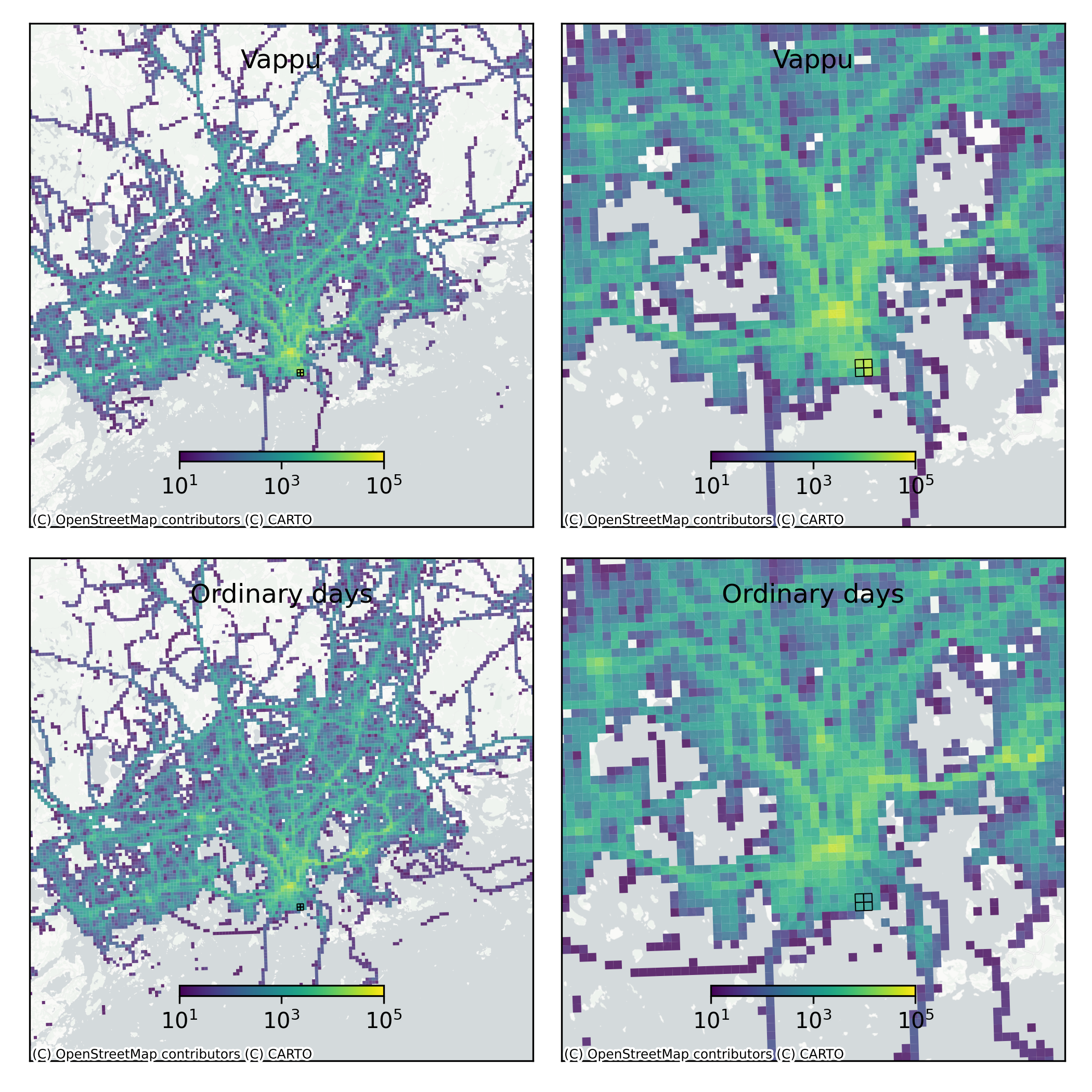}
  \caption{Comparative analysis of trajectory spatial distributions during the Vappu celebration (top row) and an ordinary non-working day (May 15th, bottom row). The right column presents a zoomed-in view ofHelsinki city center. The color gradient illustrates the number of trajectories passing through a specific grid cell.}
  \label{fig:spatial}
\end{figure}

\paragraph{Spatial distribution} Across the entire metropolitan area, the spatial distribution of trips during Vappu (Fig.\ref{fig:spatial}(a,b)) and ordinary non-working days (Fig.\ref{fig:spatial}(c,d)) exhibit similar patterns, affirming the findings of previous studies \cite{Huang2018} which point towards significant recurrent urban mobility flows. However, the spatial distribution of trips during the Vappu celebration showed a concentration of activity around the Kaivopuisto park, the primary location for Vappu celebrations (see Fig.\ref{fig:spatial}, grids in black lines). This pattern differed from ordinary Sundays when activity was more evenly distributed across the metropolitan area. Although this result was expected, since this is a known behavior for this particular day, it can serve as verification that the utilised emerging data source is capable of detecting spatial demand patterns during such events. 

\begin{figure}[tpb]
  \centering
  \includegraphics[width=8.5cm]{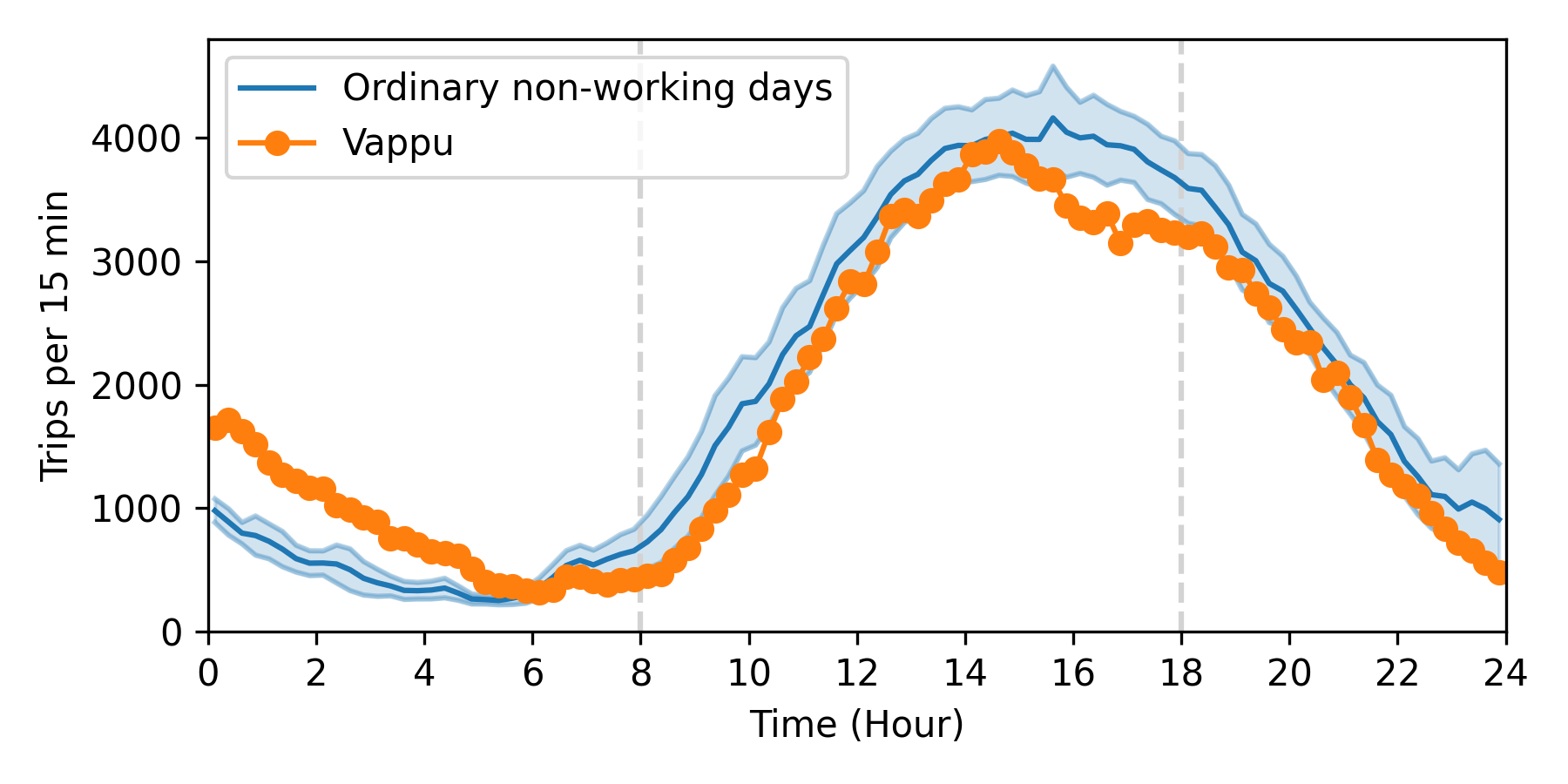}
  \caption{Comparison of the temporal distributions of trajectory numbers during Vappu (represented by the orange line) and ordinary non-working days (represented by the blue line). The upper and lower bounds of the blue line indicate one standard deviation.}
  \label{fig:temporal}
\end{figure}
    
\paragraph{Temporal distribution} Examining the temporal distribution of trip numbers across the city, the Vappu celebration period (8:00-18:00) shows slightly fewer trips than ordinary non-working days (Fig.\ref{fig:temporal}). A peak in trip volume is observed at 14:45 during Vappu, which occurs earlier compared to the average peak on ordinary non-working days. Conversely, the nighttime period (0:00-4:00) sees elevated travel volumes during Vappu. This is attributed to Vappu eve (30th April) celebrations, during which subway operations are extended until 2 a.m, May 1st.

\paragraph{Travel mode share}

According to the temporal analysis presented above, the overall travel demand of Vappu celebration remains at the same levels as ordinary non-working days (Fig.\ref{fig:temporal}). An additional research question refers to the PT mode usage and share during Vappu, focusing also on the comparison with ordinary non-working days and ordinary working days (i.e., ordinary weekdays). 

\begin{figure}[tpb]
  \centering
  \includegraphics[width=8.5cm]{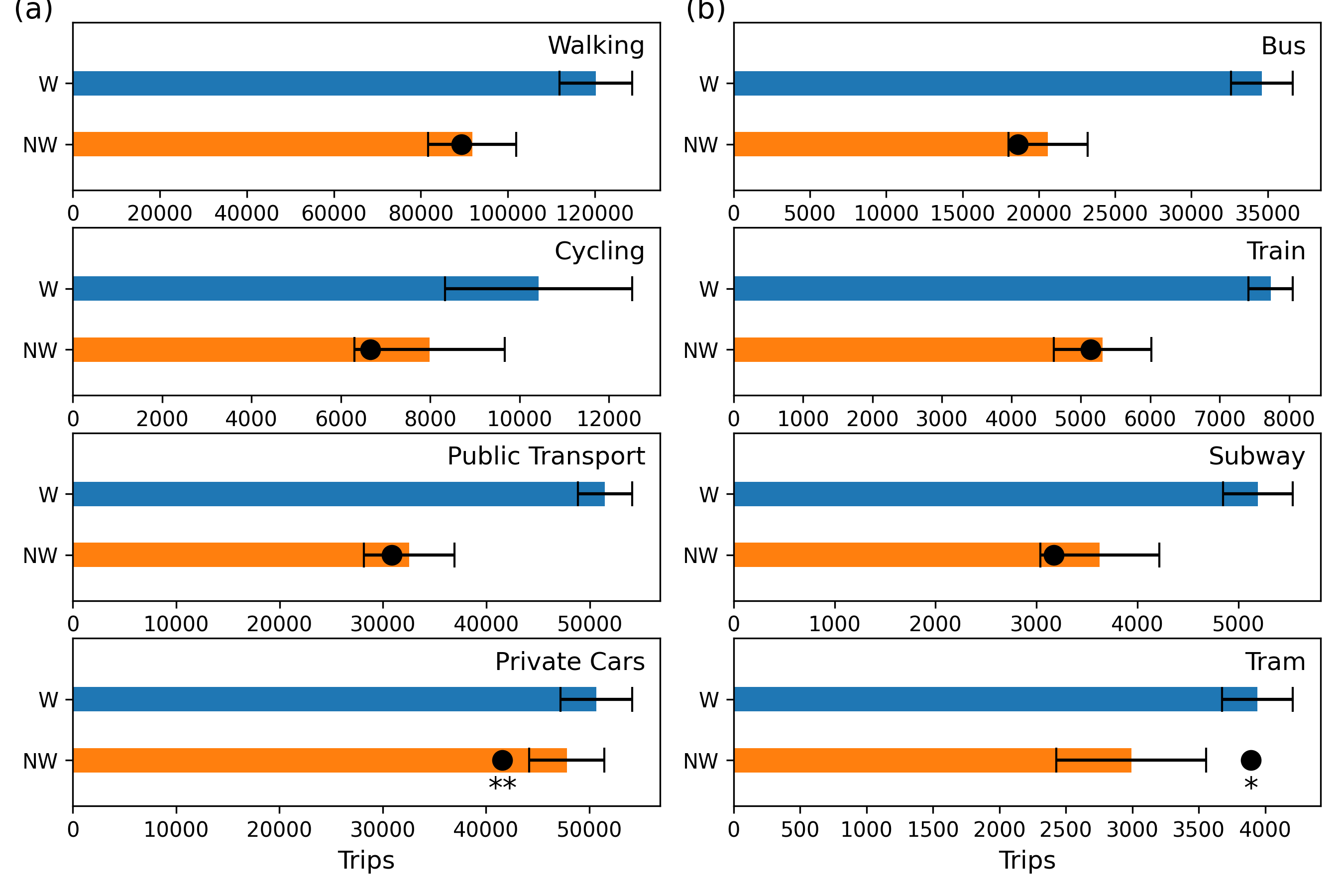}
  \caption{Trips number for different travel modes. Orange bars represent the values obtained from ordinary non-working days, blue bars represent the values obtained from working days, Black dots represent the values for Vappu, * represents the p-value of the T-test less than 0.005 and ** represents less than 0.001.}
  \label{fig:trips}
\end{figure}

\begin{figure}[tpb]
  \centering
  \includegraphics[width=8.5cm]{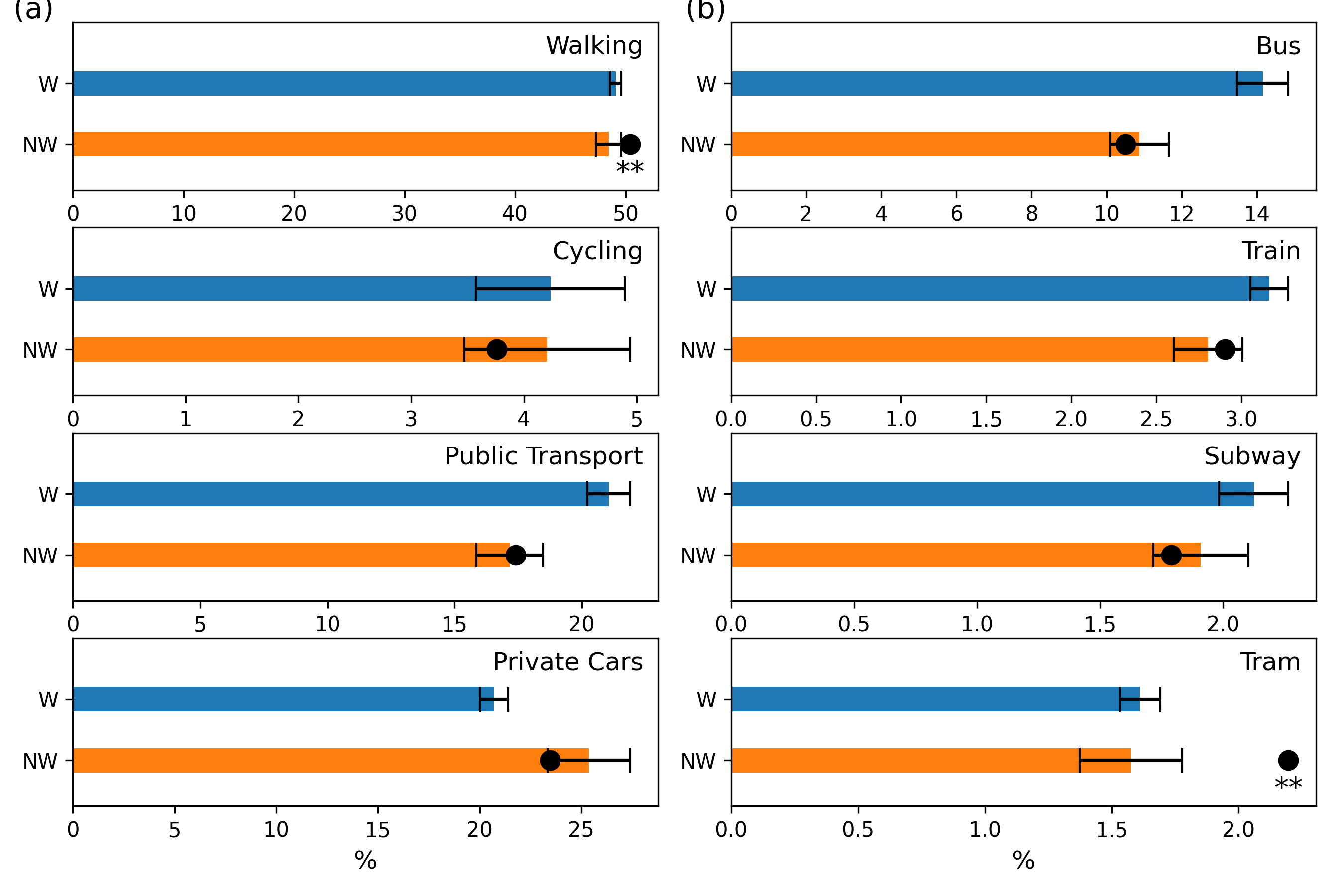}
  \caption{Percentages of total demand for different travel modes. Colours and markers are in the same format as in Fig.\ref{fig:trips}}
  \label{fig:modes}
\end{figure}

According to Fig.\ref{fig:trips}, Vappu's mode usage is within expected values for all modes, except for private cars and tram. More specifically, the usage of private cars decreases significantly during Vappu, in contrast to the usage of tram that presents a significant increase. The tram's demand is equivalent to that of an ordinary working day, while private cars users are the lowest during Vappu. These highlight some important changes that traffic and transport planners should take into account when making decisions for this special event. Fig.\ref{fig:modes} focusing on the travel mode share. According to this figure, walking presents a statistically significant increase during Vappu, the private car's share is marginally within the expected boundaries, while tram share is the greatest among the studied values.

\begin{figure}[tpb]
  \centering
  \includegraphics[width=8.5cm]{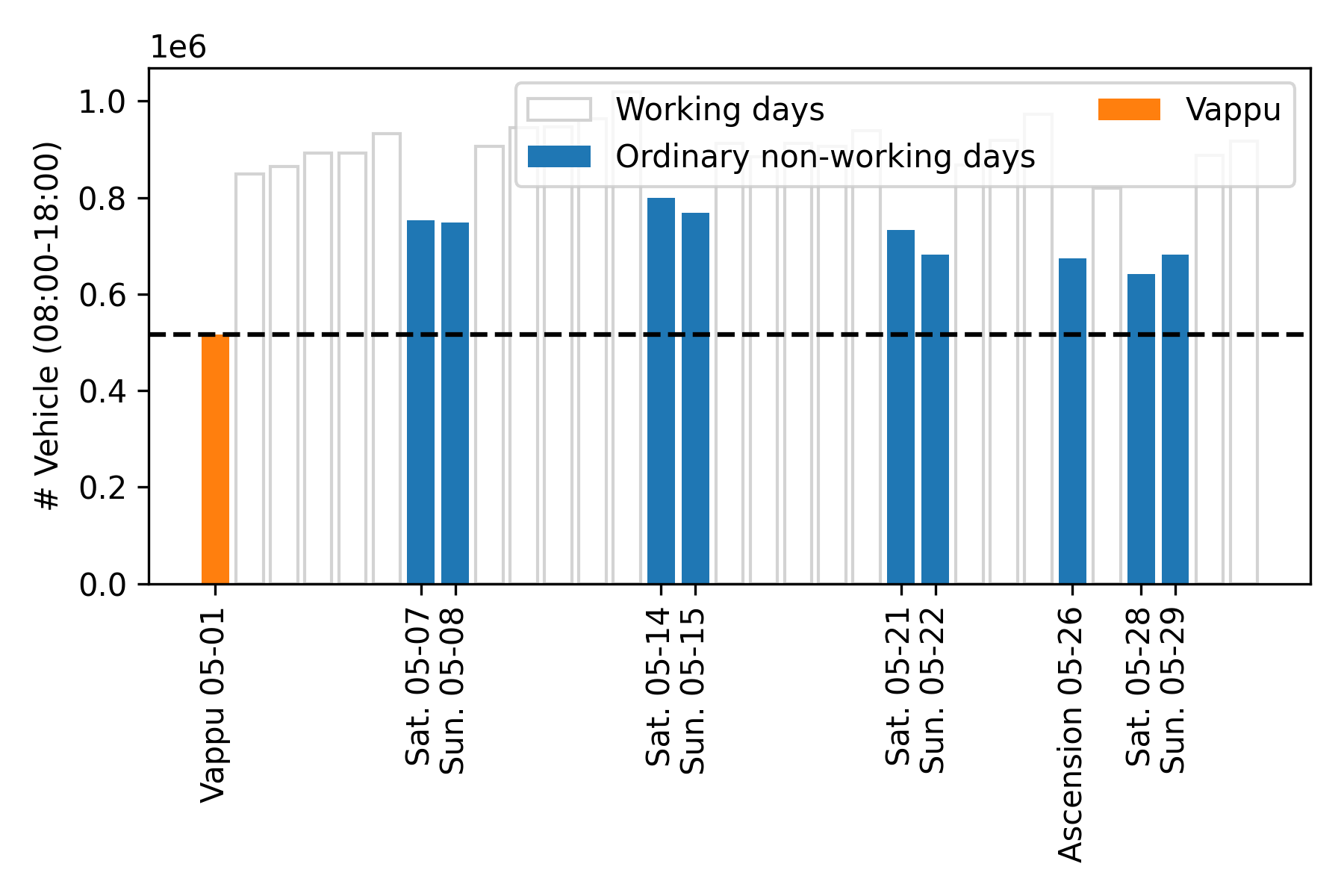}
  \caption{Private car usage through road traffic count data.}
  \label{fig:tms}
\end{figure}

For further validation of our findings, we compared the private car usage sensed by TravelSense with road traffic counts from an external data source provided by Fintraffic \cite{tms}. Fig. \ref{fig:tms} displays daily vehicle numbers, aggregated from 35 sensors situated across the capital area. It is evident that the total road traffic during Vappu is the lowest for the entire month, regardless of whether the comparison is made with working or non-working days. This observation consistently aligns with the results depicted in Fig. \ref{fig:trips}.

\subsection{Event participants' travel behaviour}

\begin{figure}[tpb]
  \centering
  \includegraphics[width=8.5cm]{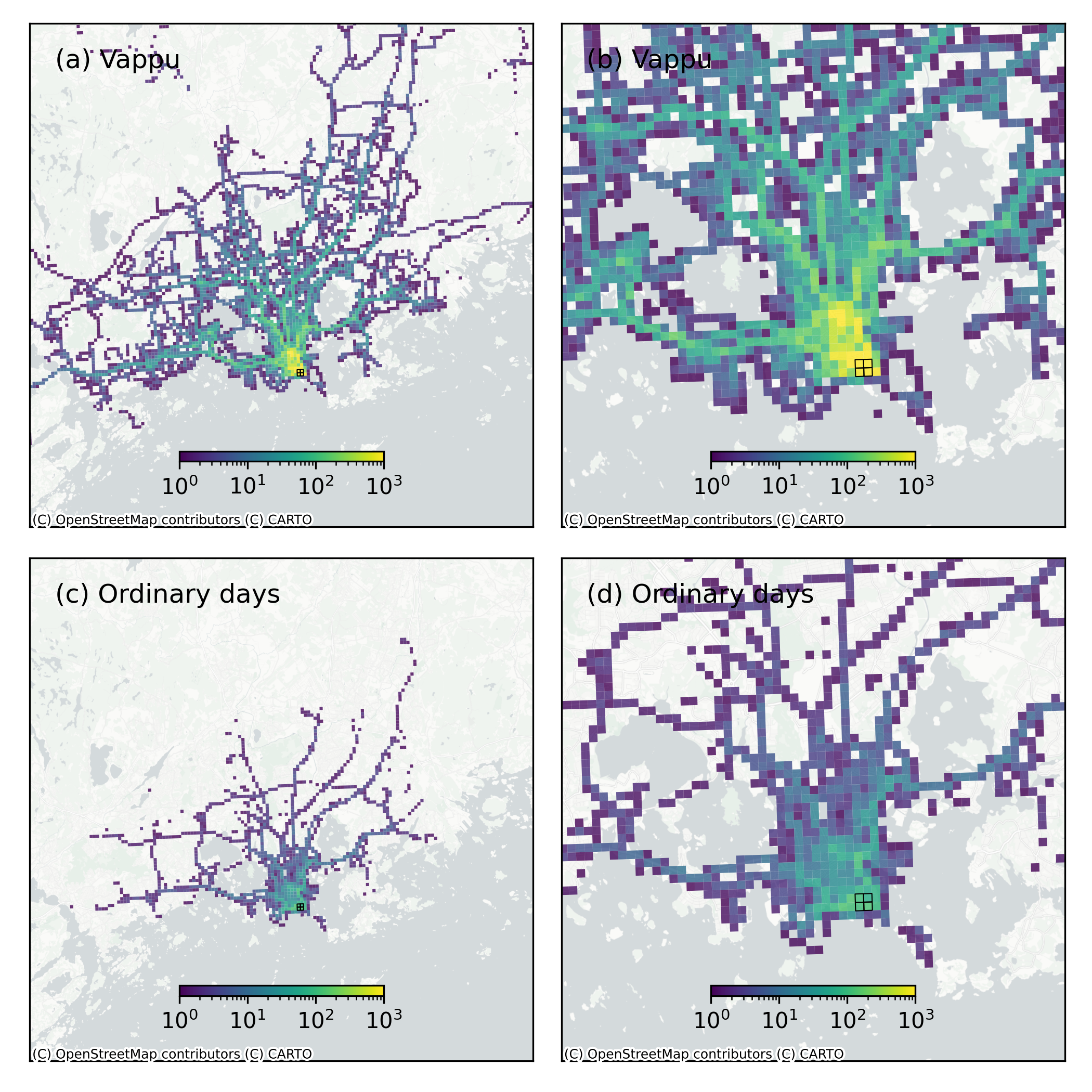}
  \caption{Heatmap of trajectories visiting the event's main location that use each grid cell during event (a, b) and during other non-working days (c, d).}
  \label{fig:traj_event}
\end{figure}

Focusing now specifically on the travelers who accessed the main event location (i.e., Kaivopuisto park), Fig.\ref{fig:traj_event} presents a heatmap resulting from the respective trajectories that visited Kaivopuisto during Vappu and other non-working days. It is apparent that the park was visited by many more travelers during Vappu compared with other days, but also that the trajectories which visited it are more scattered during the special day. It is also implied by visual inspection that the distances traveled to access the park are longer than usual.

\begin{figure}[tpb]
  \centering
  \includegraphics[width=8.1cm]{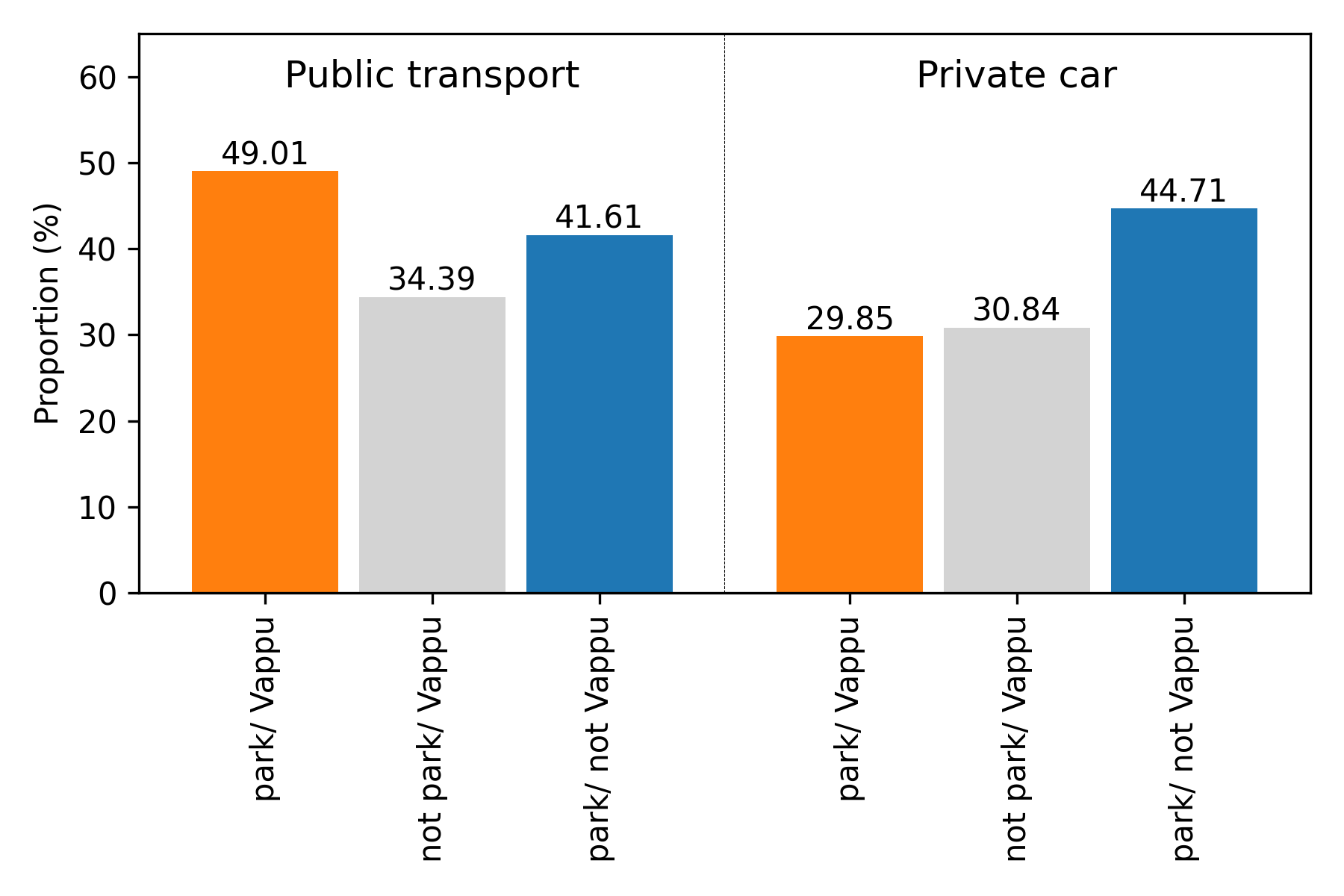}
  \caption{Proportion of trajectories that contain at least one PT trip (left) or at least one private car trip (right).}
  \label{fig:event_mode}
\end{figure}

Fig. \ref{fig:event_mode} focuses on the proportion  of trajectories that visited Kaivopuisto during Vappu (``park/ Vappu''), did not visit Kaivopuisto during Vappu (``not park/ Vappu'') and visited Kaivopuisto during ordinary non-working days (``park/ not Vappu''). The analysis refers to PT and private cars users. As shown in the figure, travelers that visited the event's main location during Vappu used PT more than private cars, in contrast to what happens during ordinary non-working days. The usage of PT among users that visited this location is also greater compared to PT users who didn't visit this location during Vappu.

\begin{figure}[tpb]
  \centering
  \includegraphics[width=8.1cm]{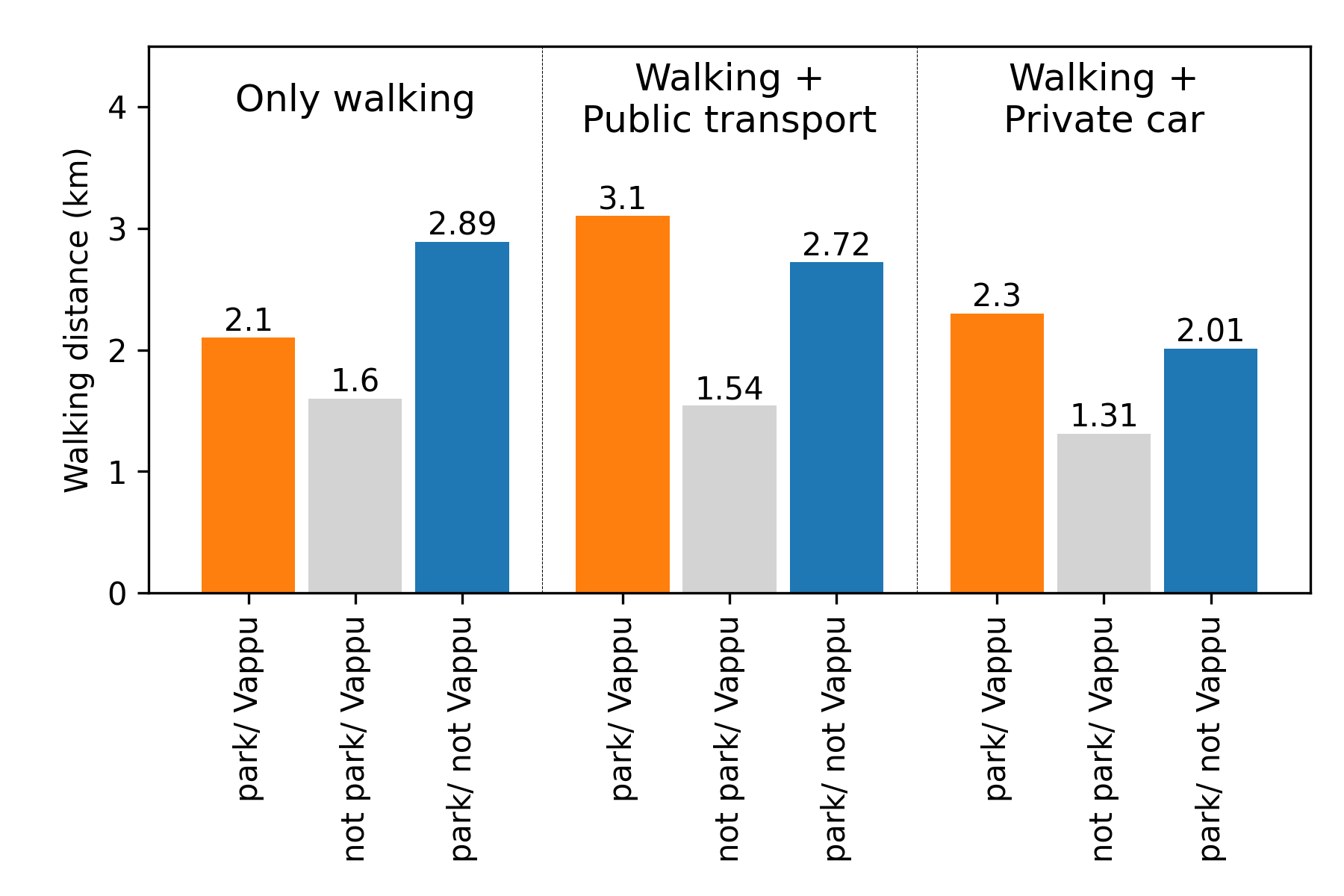}
  \caption{Walking distance for trajectories that include only walking (left), walking and PT (center), walking and private car (right).}
  \label{fig:distance}
\end{figure}

\begin{figure}[tpb]
  \centering
  \includegraphics[width=8.1cm]{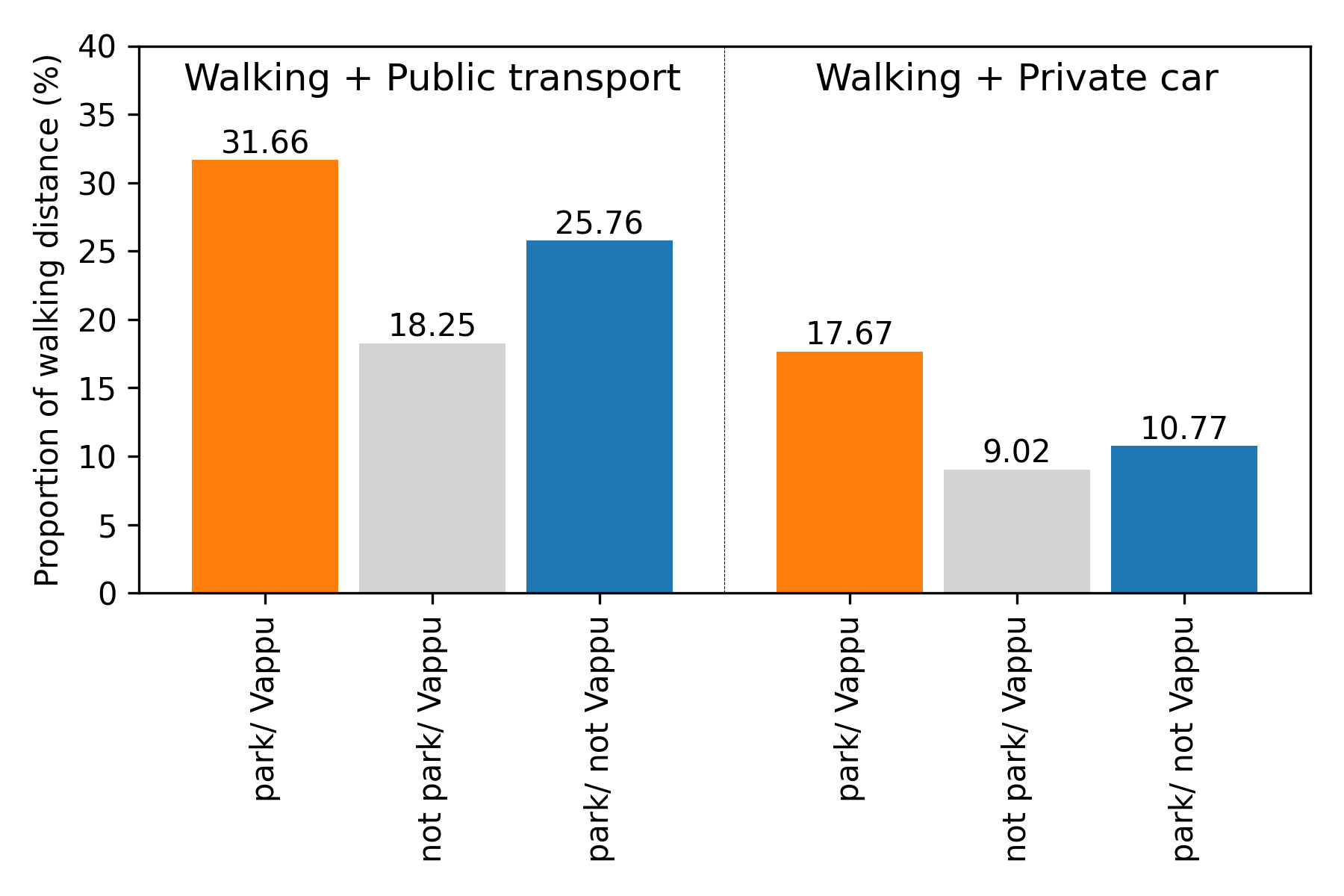}
  \caption{Proportion of total distance walked for trajectories that include walking and PT (left), walking and private car (right).}
  \label{fig:walking_share}
\end{figure}

Fig. \ref{fig:distance} shows the average walking distance for three types of trajectories, including only walking, walking and PT, walking and private cars. As shown in this figure, people are willing to walk longer distances during Vappu to join the event when the trajectory combines walking with PT or private car. When people use walking and PT as primary modes for joining the event, they are willing to walk about 3.1 km on average among the whole journey, which is the highest of three types of trajectories. Analysing the proportion of walking distance in the total distance traveled in a trajectory shows that the highest proportion of walking occurs in the trajectories that combine walking and PT (Fig.\ref{fig:walking_share}). Overall, it is observed that PT users are willing to walk more during Vappu day to access the event's main location. 

\section{Discussion and Conclusion}

This study evaluates the complex mobility patterns generated during large-scale urban events, using the Vappu celebration in Helsinki as a case study. By leveraging a unique, comprehensive dataset -- TravelSense -- from the Helsinki PT mobile application and Bluetooth beacons, we were able to capture the complexity of travel behavior during the event. Findings indicate a preference for PT over private cars and a willingness to traverse greater distances on foot for event participation, enhancing our understanding of human mobility patterns during such events.

This study offers practical implications for urban management andPT, particularly during special events. Leveraging the observed shift towards PT and increased pedestrian activity during events such as Vappu, planners and operators are positioned to enhance current strategies, accommodating for heightened demand \cite{{Currie2012}}. This may encompass the consideration of increased PT frequency, temporary route adjustments \cite{Gkiotsalitis2021}, and augmenting pedestrian facilities \cite{Eom2021}. Additionally, the data highlights the potential for promoting public and active transportation during special events, presenting an opportunity to manage congestion, reduce carbon emissions, and encourage healthier, more sustainable transport habits.

However, two principal limitations require further discussion. First, the study's findings are based on a single event, which limits the generalizability of our conclusions. Our goal here is to highlight the critical need for better understanding and planning mobility during such special events. This is particularly important for events that are repeated periodically, as the annual celebration of Vappu. Although the required efforts were more complex in the past, recent technology advancements, such as mobile data, currently allow obtaining valuable insights more easily. Second, the present analysis primarily relies on daily-based data collection, thus limiting the provision of real-time prediction or early warning capabilities. However, integrating real-time data sources and applying predictive analytics could provide early warnings \cite{Fan2015, Zhou2017, Huang2018} and proactive management options \cite{Zhou2021, Gkiotsalitis2022, Gkiotsalitis2021}. This study has the potential to contribute to the development of ITS capable of managing the complexities of large-scale urban events.

\addtolength{\textheight}{-5cm}   %

\section*{ACKNOWLEDGMENT}

The authors thank HSL (especially Pekka Räty and Hanna Kitti) for access to the data and their time for discussion. The work of Z. Huang was supported by the Strategic Research Council at the Academy of Finland (grant numbers 345188 and 345183). The work of C. Sipetas was supported by the FinEst Twins Center of Excellence (H2020 Grant 856602). The work of A. Espinosa Mireles de Villafranca was supported by the Academy of Finland. Calculations were performed using computer resources within the Aalto University School of Science “Science-IT” project.


\begin{thebibliography}{99}
\bibitem{McGillivray2019} D. McGillivray, “Sport events, space and the ‘Live City,’” Cities, vol. 85, pp. 196–202, 2019. 
\bibitem{Richards2015} G. Richards and R. Palmer, Eventful Cities. London: Routledge, 2015. 

\bibitem{Parkes2016} S. D. Parkes, A. Jopson, and G. Marsden, “Understanding travel behaviour change during mega-events: Lessons from the London 2012 games,” Transportation Research Part A: Policy and Practice, vol. 92, pp. 104–119, 2016. 
\bibitem{Guo2022} B. Guo, H. Yang, H. Zhou, Z. Huang, F. Zhang, L. Xiao, and P. Wang, “Understanding individual and collective human mobility patterns in twelve crowding events occurred in Shenzhen,” Sustainable Cities and Society, vol. 81, p. 103856, 2022. 
\bibitem{Johansson2012} A. Johansson, M. Batty, K. Hayashi, O. Al Bar, D. Marcozzi, and Z. A. Memish, “Crowd and environmental management during mass gatherings,” The Lancet Infectious Diseases, vol. 12, no. 2, pp. 150–156, 2012. 
\bibitem{Xu2017} Y. Xu and M. C. González, “Collective benefits in traffic during mega events via the use of Information Technologies,” Journal of The Royal Society Interface, vol. 14, no. 129, p. 20161041, 2017. 

\bibitem{Humphreys2018} B. R. Humphreys and H. Pyun, “Professional sporting events and traffic: Evidence from U.S. cities,” Journal of Regional Science, vol. 58, no. 5, pp. 869–886, 2018.
\bibitem{Wang2022} P. Wang, Z. Huang, J. Lai, Z. Zheng, Y. Liu, and T. Lin, “Traffic speed estimation based on multi-source GPS data and mixture model,” IEEE Transactions on Intelligent Transportation Systems, vol. 23, no. 8, pp. 10708–10720, 2022. 

\bibitem{Helbing2007} D. Helbing, A. Johansson, and H. Z. Al-Abideen, “Dynamics of crowd disasters: An empirical study,” Physical Review E, vol. 75, no. 4, 2007.
\bibitem{Pretorius2013} M. Pretorius, S. Gwynne, and E. R. Galea, “Large crowd modelling: An analysis of the duisburg love parade disaster,” Fire and Materials, vol. 39, no. 4, pp. 301–322, 2013.
\bibitem{Sharma2023} A. Sharma et al., “Global mass gathering events and deaths due to crowd surge, stampedes, Crush and physical injuries – lessons from the Seoul Halloween and other disasters,” Travel Medicine and Infectious Disease, vol. 52, p. 102524, 2023.  

\bibitem{Kaiser2018} M. S. Kaiser, K. T. Lwin, M. Mahmud, D. Hajializadeh, T. Chaipimonplin, A. Sarhan, and M. A. Hossain, “Advances in crowd analysis for urban applications through urban event detection,” IEEE Transactions on Intelligent Transportation Systems, vol. 19, no. 10, pp. 3092–3112, 2018.  

\bibitem{Candia2008} J. Candia et al., “Uncovering individual and collective human dynamics from Mobile Phone Records,” Journal of Physics A: Mathematical and Theoretical, vol. 41, no. 22, p. 224015, 2008.
\bibitem{Dong2015} Y. Dong, F. Pinelli, Y. Gkoufas, Z. Nabi, F. Calabrese, and N. V. Chawla, “Inferring unusual crowd events from mobile phone call detail records,” Machine Learning and Knowledge Discovery in Databases, pp. 474–492, 2015. 
\bibitem{Fan2015} Z. Fan, X. Song, R. Shibasaki, and R. Adachi, “Citymomentum: an online approach for crowd behavior prediction at a citywide level,” Proceedings of the 2015 ACM International Joint Conference on Pervasive and Ubiquitous Computing - UbiComp '15, 2015. 
\bibitem{Zhou2017} J. Zhou, H. Pei, and H. Wu, “Early warning of human crowds based on query data from Baidu Maps: Analysis based on Shanghai stampede,” Advances in Geographic Information Science, pp. 19–41, 2017. 

\bibitem{Campanella2013}  M. Campanella, R. Halliday, S. Hoogendoorn, and W. Daamen, “Managing large flows in metro stations: Lessons learned from the New Year celebration in Copacabana,” 16th International IEEE Conference on Intelligent Transportation Systems (ITSC 2013), 2013. 

\bibitem{Huang2018} Z. Huang, P. Wang, F. Zhang, J. Gao, and M. Schich, “A mobility network approach to identify and anticipate large crowd gatherings,” Transportation Research Part B: Methodological, vol. 114, pp. 147–170, 2018. 

\bibitem{Batty2012} M. Batty, J. Desyllas, and E. Duxbury, “Safety in numbers? modelling crowds and designing control for the Notting Hill Carnival,” Urban Studies, vol. 40, no. 8, pp. 1573–1590, 2003.
\bibitem{Klüpfel2013} H. Klüpfel and S. Hebben, “Large scale outdoor events: Specific Requirements Concerning Evacuation Analysis,” Pedestrian and Evacuation Dynamics 2012, pp. 501–508, 2013. 
\bibitem{Zhou2021} H. Zhou, Z. Zheng, X. Cen, Z. Huang, and P. Wang, “A data-driven urban metro management approach for crowd density control,” Journal of Advanced Transportation, vol. 2021, pp. 1–14, 2021. 
\bibitem{Gkiotsalitis2022} K. Gkiotsalitis, “A dynamic stop-skipping model for preventing public transport overcrowding beyond the pandemic-imposed capacity levels,” 2022 IEEE 25th International Conference on Intelligent Transportation Systems (ITSC), 2022. 

\bibitem{hsl} “Journey Planner, tickets and fares, Customer Service,” HSL.fi, https://www.hsl.fi/en (accessed May 21, 2023). 
\bibitem{Huang2022} Z. Huang, A. Espinosa Mireles de Villafranca, and C. Sipetas, “Sensing Multi-modal Mobility Patterns: A Case Study of Helsinki using Bluetooth Beacons and a Mobile Application,” in Proceedings of the 2022 IEEE International Conference on Big Data (Big Data). Osaka, Japan: IEEE, pp. 2007–2016. 2022.
\bibitem{Raty2018} Brandt, E., Kantele, S. and Räty, P. “Liikkumistottumukset Helsingin seudulla 2018 (Travel habits in the Helsinki region in 2018),” Helsinki: HSL Helsinki Region Transport, 2019.
\bibitem{Rinne2017} M. Rinne, M. Bagheri, T. Tolvanen and J. Hollmén, "Automatic Recognition of Public Transport Trips from Mobile Device Sensor Data and Transport Infrastructure Information", Personal Analytics and Privacy. An Individual and Collective Perspective, pp. 76-97, 2017.
\bibitem{Huang2023} Z. Huang, A. Espinosa Mireles de Villafranca, and C. Sipetas, and T. Quach, “Crowd-sensing commuting patterns using multi-source wireless data: a case of Helsinki commuter trains” 2023. [Online]. Available: arXiv:2302.02661.
\bibitem{Yle2022} “Kaivopuistossa Nautittiin Vappupiknik, Helsingissä Marssittiin, Osalla Poliiseista kiireinen viikonloppu – Seurasimme Vapun Viettoa Koko Maassa,” Yle Uutiset, https://yle.fi/a/3-12424506 (accessed May 16, 2023). 
\bibitem{tms} “TMS documentation,” Digitraffic, https://www.digitraffic.fi/en/road-traffic/lam/ (accessed May 16, 2023). 
\bibitem{Currie2012} G. Currie and A. Shalaby, “Synthesis of transport planning approaches for the world’s largest events,” Transport Reviews, vol. 32, no. 1, pp. 113–136, 2012.
\bibitem{Gkiotsalitis2021} K. Gkiotsalitis and O. Cats, “At-stop control measures in public transport: Literature Review and research agenda,” Transportation Research Part E: Logistics and Transportation Review, vol. 145, p. 102176, 2021.
\bibitem{Eom2021} S. Eom and Y. Nishihori, “How weather and special events affect pedestrian activities: Volume, space, and Time,” International Journal of Sustainable Transportation, vol. 16, no. 5, pp. 462–475, 2021.


\end{thebibliography}
\end{document}